\documentclass[pra,twocolumn,eqsecnum,superscriptaddress,showpacs,amsmath,amstex,amssymb,citeautoscript]{revtex4-1}

\usepackage[T1]{fontenc}
\usepackage[utf8]{inputenc}
\usepackage{lipsum}
\usepackage{amsmath}
\usepackage{amssymb}
\usepackage{bbm,bm}
\usepackage{braket}
\usepackage{xcolor}
\usepackage{pifont}
\usepackage[mathscr]{euscript}
\usepackage[shortlabels]{enumitem}
\usepackage[justification=justified]{subcaption}
\usepackage{graphicx}
\usepackage{lipsum}
\allowdisplaybreaks
\usepackage{float}
\usepackage{graphicx}
\usepackage{dsfont}
\usepackage{comment}
 
\usepackage{microtype}
\definecolor{webblue}{rgb}{0, 0, 0.5} 

\usepackage[bookmarks=true,
bookmarksnumbered=false, 
bookmarksopen=false, 
colorlinks=true,
linkcolor=webblue]{hyperref}

\hypersetup{colorlinks=true,urlcolor=webblue,citecolor=blue}

\usepackage{cellspace}
\renewcommand{\approx}{\simeq}

\begin{document}

\footnotetext[4]{In fact, recent avoided level crossing $\mu$SR has suggested there is minimal change in the relaxation rate near $T_{CDW}$, with a significant signal near $T_*$ rather than $T_{CDW}$ \cite{bonfa2024unveiling}.}

\footnotetext[5]{The switching of the Bragg peaks has since been reproduced in a recent experiment \cite{piezo_xing2024}.}

\footnotetext[6]{The possibility of a CDW state with domains whose magnetic field or $C_2$ orientation average to zero is highly unlikely as an explanation for the consistent and reproducible absence of $C_3$ symmetry breaking observed in low-strain samples. The results of Ref. \cite{anisotropic_Guo2024} were demonstrated in flakes of dimension $\sim 10-20$ microns, which would require domains of dimension $\sim$ nm for such a cancellation. Setting aside the energetic cost of such nanoscale variations in the charge ordering, there is no evidence for domains of this scale in local probes \cite{Xu2022, zhao2021cascade}.}

\footnotetext[7]{J. Lin, private communication.}

\footnotetext[8]{R. Comin, unpublished.}

\footnotetext[9]{Here we have only described the Raman process involving a single phonon scattering off a static $E_{1u}$ condensate; we have neglected processes involving fluctuations of the $E_{1u}$ order.  We leave it for future work to examine the full Raman spectrum induced by $E_{1u}$ order and fluctuations thereof.}

\footnotetext[10]{Here we use point group notation to describe the CDW symmetries; in space group notation, $A_u$ and $B_u$ correspond to $M_1^-$ and $M_2^-$ respectively.}



\title{Vestigial Order from an Excitonic Mother State  in Kagome Superconductors $A$V$_3$Sb$_5$}

\author{Julian Ingham}
\affiliation{Department of Physics, Columbia University, New York, NY, 10027, USA}

\author{Ronny Thomale}
\affiliation{Institute for Theoretical Physics and Astrophysics, University of Würzburg, D-97074 Würzburg, Germany}

\author{Harley D.~Scammell}
\email[]{harley.scammell@uts.edu.au}
\affiliation{School of Mathematical and Physical Sciences, University of Technology Sydney, Ultimo, NSW 2007, Australia}

\date{\today}

\begin{abstract}
    
    Alongside high-temperature charge order and superconductivity, kagome metals exhibit signatures of time-reversal symmetry breaking and nematicity which appear to depend strongly on external perturbations such as strain and magnetic fields, posing a fundamental challenge for conceptual reconciliation. We develop a theory of vestigial order descending from an excitonic mother state in $A$V$_3$Sb$_5$ ($A$=K,Rb,Cs), which develops around $T_* \approx 40$ K. The application of external fields stabilises a subset of the phase-melted order parameter manifold, referred to as a vestigial state, producing a symmetry-breaking response which depends on the applied probe. Our theory reproduces the observations of piezomagnetism, electric magnetic chiral anisotropy, absence of Kerr rotation, unusual elastoresistance response, and superconducting diode effect. Our proposed excitonic mother state accounts for probe-dependent symmetry breaking patterns without fine-tuning, and predicts additional signatures accessible through optical spectroscopy.
\end{abstract}

\maketitle


\section{Introduction} 
Kagome metals have emerged as a rich playground for the exploration of correlated phases of matter, and developed into one of the most vibrant fields in contemporary condensed matter physics due to the propensity of these systems to host exotic electronic orders~\cite{Tan2011, Green2010, yu2012chiral, Kiesel2013, Kiesel2012, Li2022, wenger2024theory, Scammell2023, ingham2024theory, dong2023loop, profe2024kagome, nag2024pomeranchuk, wang2025formation, jiang2024van, wang2023quantum, yin2022topological, neupert2022review}. The recently discovered superconducting family $A$V$_3$Sb$_5$ ($A$=K,Rb,Cs) has attracted immense attention due to its exotic principal phenomenology of high-temperature exotic charge order which sets in around $T_{CDW}\approx 100$ K, coexisting with  unconventional superconductivity with $T_c\approx 2$ K~\cite{PhysRevMaterials.3.094407, PhysRevLett.125.247002, yang2021giant, Chen2022anomalous, Mielke2021b, jiang2021np, Xu2022, li2021rotation, saykin2023high, zhao2021cascade, Xu2021, Gupta2021, Duan2021, Li2021c, Shumiya2021, Miao2021, Ni2021, Zhu2021, Chen2021b, Du2021b, Zhang2021, Liang2021, ortiz2021fermi, oey2022fermi, kang2023charge, huang2025revealing, hossain2025field}. As theoretical proposals continue to proliferate  exciting predictions -- such as orbital current phases, nematic orders, and pair density wave superconductivity~\cite{PhysRevLett.127.177001, PhysRevLett.127.217601, PhysRevB.110.024501, PhysRevB.108.L081117, christensen2021theory, christensen2022loop, zhou2022chern, li2024intertwined, tazai2022mechanism, tazai2023charge, tazai2024drastic, tazai2024quantum, feng2025odd} -- the accumulating experimental phenomenology from multiple measurement techniques such as surface spectroscopy, transport, and thermodynamics has proven as intriguing as it is difficult to reconcile within a universal framework \cite{wilson2024kagome}.

This particularly applies to the intermediate regime $T_c<T<T_{CDW}$, in which experiments have presented seemingly contradictory results. A recurring question has been whether the CDW appearing at $T_{CDW}$ breaks rotational or time-reversal symmetry (TRS). Initial reports claimed signatures of TRS breaking at $T_{CDW}$ -- pointing to a large Hall conductivity \cite{yang2021giant}, an increase in the relaxation rate in $\mu$SR \cite{Mielke2021b,jiang2021np}, magnetic field-tuned switching of the CDW Bragg peaks \cite{jiang2021np}, and Kerr rotation \cite{Xu2022}. Yet the experimental status has significantly complicated since these initial reports -- no hysteresis is seen in the Hall conductivity \cite{yang2021giant, Chen2022anomalous}, many experiments see no switching of the Bragg peaks \cite{li2021rotation, Note5} or Kerr rotation \cite{saykin2023high}, and while $\mu$SR does see a feature near $T_{CDW}$, the largest increase in relaxation rate is observed near $T_{*} \approx 40$ K \cite{Note4}. 

Recent experiments suggest a clarifying perspective on these unusual properties. When placed in an ultra low-strain environment, CsV$_3$Sb$_5$ shows evidence of inversion symmetry breaking in nonlinear transport near $T_*$, known as electric-magnetic chiral anisotropy (eMChA), while these signatures rapidly vanish in small strain fields \cite{emcha_Guo2024}. Magnetic torque measurements also observe thermodynamic evidence of a magnetic transition around $T_*$ \cite{Note7}, at which a large and unusual elastoresistance response in the symmetric $A_{1g}$ channel is observed \cite{elasto_liu2023}. Also near $T_*$, a giant piezomagnetic response appears \cite{piezo_xing2024}, and small out-of-plane magnetic fields result in threefold rotational symmetry breaking \cite{anisotropic_Guo2024}. Of particular note is the probe-dependence of the symmetry breaking – $c$-axis strain or out-of-plane magnetic fields eliminate eMChA \cite{Guo2022}, while low-strain samples \textit{only} present signatures of rotational symmetry breaking in applied magnetic fields \cite{anisotropic_Guo2024, Note6}. Taken together, these experiments demonstrate that in the absence of perturbations, CsV$_3$Sb$_5$ does not break TRS nor rotational symmetries in the absence of strain or magnetic fields, yet develops some kind of unusual electronic phase near $T_*$ that exhibits a drastic symmetry breaking response to external probes. This large inherent sensitivity of the broken symmetries to external conditions rationalises the contradictory results from different experimental groups.

\begin{figure*}[t]
    \centering
    \includegraphics[width=\textwidth]{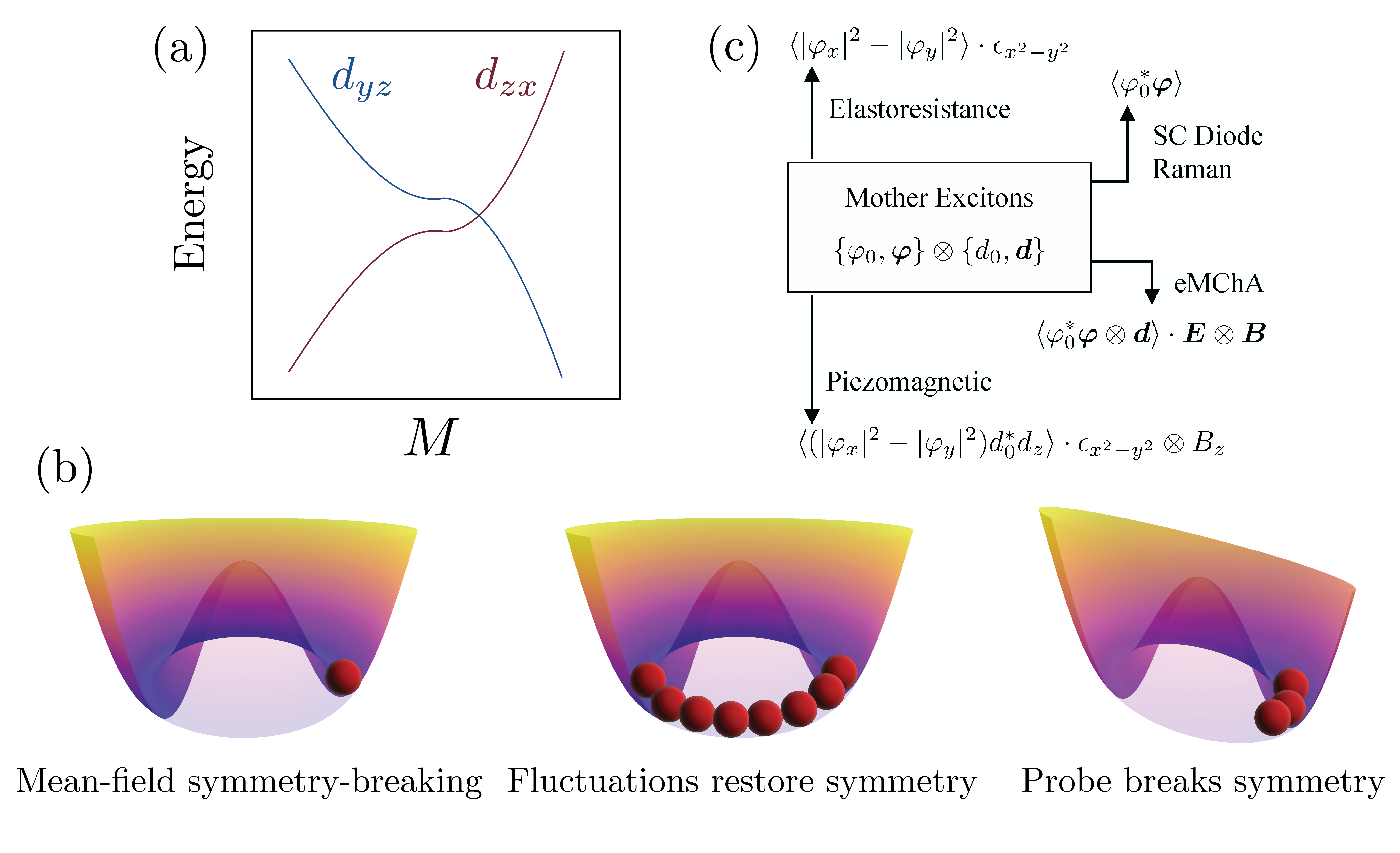}
    \vspace{-0.6cm}
    \caption{\textbf{Probe-dependent symmetry breaking from an excitonic mother state.} (a) The bandstructure of $A$V$_3$Sb$_5$ features twofold van Hove singularities near the $M$ point with opposite concavity, originating from the $d_{xz}$ and $d_{yz}$ vanadium orbitals. These bands are prone towards an excitonic instability which hybridises the bands. (b) Concept of a vestigial mother state: an excitonic order parameter breaks a large set of symmetries at mean-field level, represented by the red ball in a minimum of the wine bottle potential, but fluctuations between different symmetry-related ground states restore the symmetry on average. Applying an external probe quenches these fluctuations, resulting in a large symmetry-breaking response, represented by a strong accumulation of spectral weight in a symmetry-breaking minimum of the free energy. (c) Different external probes -- applied strain, electric and magnetic fields -- stabilise different vestigial orders, resulting in a range of unconventional responses.}
    \label{fig1}
    \vspace{-0.3cm}
\end{figure*}

From a synoptic experimental perspective on kagome superconductors, this presents an exciting puzzle: is it possible to capture all such features through a single organisational principle? The outset might bear some similarities to the pseudogap phase in high-temperature copper oxide superconductors, where the system's response likewise sensitively depends on the precise parameters and external conditions \cite{keimer2015quantum}. Cuprates witnessed the pseudogap phase as an intricate interpolation phenomenon between the dominant magnetic phase at half filling and the unconventional superconducting phase at sufficient doping; in this context the concept of intertwined orders has arisen, via which the comparable multiple ordering propensities of a correlated electronic system are more naturally explained than from merely competing orders~\cite{RevModPhys.87.457}. Cuprates further suggest a pair density wave as potential ``mother description'', out of which the charge orders derive as vestigial states~\cite{annurev:/content/journals/10.1146/annurev-conmatphys-031119-050711}.  By analogy, the key ordering pillars of kagome metals would be charge order and superconductivity. Those ordered domains are, however, much less well-understood than for the high-$T_c$ cuprates, and exhibit a totally unique set of symmetry-breaking patterns.

Existing theories have attempted to rationalize the experimental phenomenology of kagome metals through multiple CDW orders appearing simultaneously at $T_*$ \cite{piezo_xing2024, WagnerPRB2023, tazai2024quantum}. Such ansätze are somewhat contrived due to the implicit fine tuning in positing multiple coincident order parameters. Instead, we propose a different conceptual framework to unify these observations: that of a new phase of matter, a vestigial excitonic state. The bandstructure of kagome metals such as $A$V$_3$Sb$_5$ features electron- and hole-like bands close to the Fermi level. From there, it has been proposed that $d$-wave excitonic order is a natural ground state for repulsive interactions, and can coexist with CDW order \cite{Scammell2023}. In a quasi-two-dimensional system, such a state could serve as a ``mother state'' that is melted by phase fluctuations. We show that external probes such as field or strain stabilise bilinears of this melted order parameter, i.e. form vestigial orders \cite{nie2014quenched, fernandes2019intertwined, berg2009charge, poduval2024vestigial}. The symmetry of the vestigial order depends on the applied probe (Fig. \ref{fig1}b); here we show that the richness of the spin and orbital structure of this melted order parameter manifold naturally synthesises a large range of experimental observations, presenting a unified theory for kagome superconductors such as CsV$_3$Sb$_5$. Building on this synthesis, we predict signatures of our putative mother state in optical spectroscopy and STM.

By comparison to the pair density wave as a tentative parent state of high-$T_c$ cuprates, our proposal can be interpreted as a description for kagome metals hinging on excitonic fluctuations inherited from a melted excitonic mother state.

\section{Results}

\subsection{Parent excitonic order}

We begin with a discussion of the orbital content of the bands in $A$V$_3$Sb$_5$. ARPES observes two oppositely dispersing bands near the Fermi level, which arise from the multiplet of vanadium $d$-orbitals \cite{Kang2022, Hu2022}; as illustrated schematically in Fig. \ref{fig1}a. The minimal two-band model we employ comprises $d_{zx}, d_{yz}$ orbitals -- for more discussion of the bandstructure and orbital content see the Supplementary Material (SM).   Describing the vHS nearest to the Fermi level (Fig. \ref{fig1}), the single particle Hamiltonian is given by
\begin{align}
\label{H0}
    H_0 & = \sum_{\bm k} c^\dag_{\bm k} (\varepsilon^c_{\bm k}-\mu) c_{\bm k} + v^\dag_{\bm k} (\varepsilon^v_{\bm k} - \mu) v_{\bm k}
\end{align}
where $c^\dag_{\bm k}$, $(v^\dag_{\bm k})$ are creation operators for conduction (valence) electrons, with spin indices suppressed. The U(1) symmetry manifests as independent global U(1) phases of the $c^\dag_{\bm k}, v^\dag_{\bm k}$ electrons; in the limit where this symmetry is exact, the conduction and valence bands correspond to pure $d_{zx}$ and $d_{yz}$ content. The chemical potential is tuned to the approximate vicinity of the $M$-points, near which there is an approximate spectral particle-hole symmetry $\varepsilon^v_{\bm M_i+\bm k}\approx -\varepsilon^c_{\bm M_i+\bm k}$. Finally, we make special note that, at the $M$-points, $c^\dag_{\bm k}$ ($v^\dag_{\bm k}$) create $p$-type ($m$-type) vHS states \cite{Kiesel2012}; these are inversion even (odd).

The opposite dispersion of these two bands results in an instability towards an excitonic condensate, as theoretically conjectured in Ref. \cite{Scammell2023}, where a tendency towards $d$-wave excitonic pairing -- i.e. particle-hole pairing in a two-dimensional (2D) irreducible symmetry representation (irrep) -- is found. We take this premise as a starting point for the present work, and elevate it to form a basis for our theory of a vestgial mother state of kagome metals.

Utilizing the results of Ref. \cite{Scammell2023}, we specialize to pairing in a 2D irrep.  The pairing term in the Hamiltonian takes the form 
\begin{align}
\label{Hpair}
    H_\text{pair} & = \sum_{\bm k}  M_{\bm k} \, c^\dag_{\bm k}  v_{\bm k}  +\text{h.c.}, \
    M_{\bm k} = \left(\bm\varphi \cdot {\cal E}_{\bm k}\right)d_\mu s_\mu
\end{align}
where $s_\mu$ ($\mu=0,x,y,z$) are Pauli matrices acting on spin and ${\cal E}_{\bm k}$ is a two-component basis function that represents the different possible 2D irreps. Combining states from vHS of opposite parity, p-type/m-type, and with a $d$-wave pairing, as per \cite{Scammell2023}, produces an order parameter which transforms overall as the $E_{1u}$ irrep \footnote{As an alternative to this microscopic viewpoint, one may consider particle-hole pairing in the $E_{1u}$ irrep as a postulate of the present work.}, which we parametrize as
\begin{align}
   \notag {\cal E}_{\bm k}^{E_{1u}} &= (X_{\bm k}, Y_{\bm k}) \in E_{1u}
\end{align}
where $X_{\bm k}, Y_{\bm k}$ transform as $k_x, k_y$. The order parameter in Eq. \eqref{Hpair} is composed of two complex vectors 
\begin{align}
    \bm\varphi &= (\varphi_x,\varphi_y), \
    \bar{d} = (d_0,d_x,d_y,d_z).
\end{align}
The basis function $\bm\varphi\in E_{1u}$ transforms as a vector under spatial rotations. The three-vector $\bm d =(d_x,d_y,d_z)$ transforms as an SO(3) pseudo-vector under spin-rotations and corresponds physically to spin-triplet pairing, while $d_0$ is a scalar and corresponds to spin-singlet pairing. We will denote the order parameter as the tensor product of the spatial and spin components $\Delta=\bm\varphi \otimes \bar{d}$. 

In the absence of exchange interactions, there is separate conservation of the spin for each flavour, i.e. spin SU(2)$\times$SU(2) symmetry -- which manifests as an SO(4) spin symmetry of the excitonic order parameters, implying that spin-singlet and triplet states form a degenerate manifold. Including exchange terms, the order parameter decomposes into singlet and triplet channels. Our conclusions do not rely on strict SO(4) symmetry; in the SM we allow for perturbations away from this limit.

Finally, previous {\it ab initio} studies  show explicit breaking of orbital U(1) symmetry \cite{PhysRevLett.127.177001}. To account for this, we additionally introduce a U(1)-breaking background field term $\varphi_0 \chi^{B_{2u}}_{\bm k}c^\dag_{\bm k} v_{\bm k}$ into the pairing Hamiltonian \eqref{Hpair}, where $\varphi_0$ is a spin-independent complex number with phase set by the microscopic details of the orbitals, while $\chi^{B_{2u}}_{\bm k}$ is a basis function in the $B_{2u}$ irrep. Crucially, this background field breaks the global U(1) but does not break crystalline point group $D_{6h}$, i.e. transforms as the $A_{1g}$ irrep.  The background field $\varphi_0$ allows for the $\Phi_{E_{1u}}$ vestigial of Table \ref{tab:vestigials}.

\setlength{\tabcolsep}{10pt} 
\renewcommand{\arraystretch}{1.75} 
\begin{table}[b]
    \centering
    \begin{tabular}{|c|c|c|c|}
        \hline \hline
        \multicolumn{2}{|c|}{Spatial} & \multicolumn{2}{|c|}{Spin} \\ \hline
        Label & Content & Label & Content \\ \hline
        $\Phi_{A_{1g}}$ & $\bm{\varphi}^*\cdot\bm{\varphi}$ & $D_\text{scalar}$ & $\bar{d}^*\cdot \bar{d}$ \\ \hline
        $\Phi_{A_{2g}}$ & $i\bm{\varphi}^*\times\bm{\varphi}$ & $\bm D_\text{axial}$ & $d_0^*\bm{d}$ \\ \hline
        $\bm \Phi_{E_{2g}}$ & $\begin{pmatrix} \varphi_x^*\varphi_x-\varphi_y^*\varphi_y\\ \varphi_x^*\varphi_y+\varphi_y^*\varphi_x \end{pmatrix}$ & $\tilde{\bm D}_\text{axial}$ & $\bm{d}^*\times \bm{d}$ \\ \hline
        $\bm \Phi_{E_{1u}}$ & $\varphi_0^*\bm{\varphi}$ & - & - \\ \hline
    \end{tabular}
    \caption{Possible vestigial orders decomposed into spatial and spin contributions $\Phi_\Gamma \otimes D_{\Gamma_s}$. }
    \label{tab:vestigials}
\end{table}
\subsection{Vestigial order}

\begin{figure*}[t!]
  \begin{center}
\includegraphics[width=\textwidth]{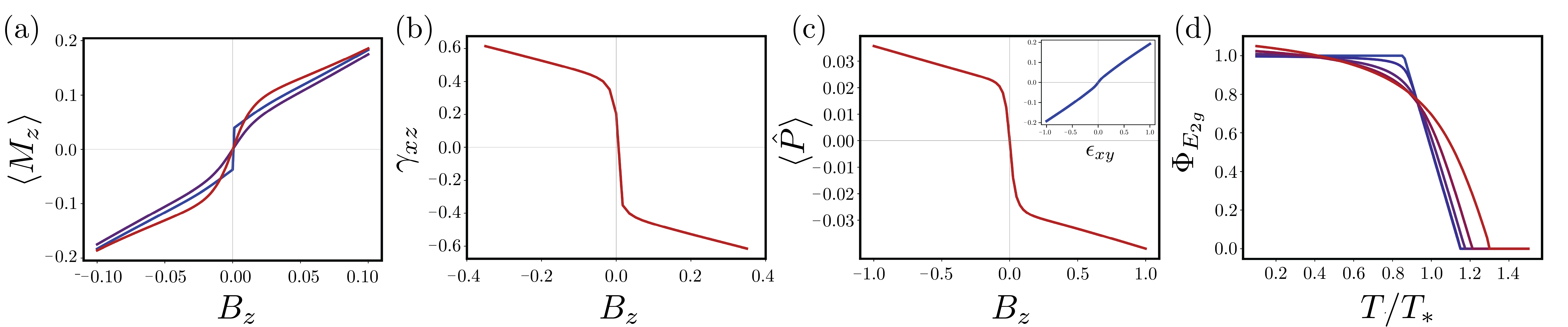}
    \caption{\textbf{Unconventional responses of the excitonic mother state.} (a) AHE with increasing values of strain from blue to red, as measured by the magnetization $\braket{M_z}$. (b) The eMChA response, defined so that the resistance receives a contribution $\gamma_{ij}B_iI_j$. (c) Piezomagnetic response, measured by the expectation value  of $\hat{P}=\Phi_{E_{2g}}\otimes D''_{\text{axial}}$, which requires both non zero field and strain. Red shows the dependence of $\hat{P}$ on field for fixed strain, and blue the dependence on strain for fixed $B_z\neq0$. (d) Transport anisotropy measured by the magnitude of the order parameter $\Phi_{E_{2g}}$, for incrementally increasing $B_z$. The free energy and explicit values of the parameters employed are presented in the SM.}
\label{fig2}
	\end{center}
 \vspace{-0.6cm}
\end{figure*}

The excitonic state breaks a set of continuous approximate symmetries -- the SO(4) arising from  near degeneracy of singlet and triplet order, along with the approximate U(1) particle-hole number symmetry, and so in two dimensions this order parameter should in fact melt due to fluctuations, $\braket{\Delta}\approx0$. The importance of fluctuations to the physics of $T_*$ is underscored by the apparent broadness of the transition, seen for in e.g. piezomagnetism \cite{anisotropic_Guo2024}. On the other hand, vestigial orders -- products of different components of the order parameter manifold we denote by $V[\Delta^*,\Delta]$ -- can be invariant under the continuous symmetries, breaking only discrete ones \cite{fernandes2019intertwined}. These states are stable against fluctuation effects.

We neglect spin-orbit coupling, and thereby treat spin and spatial contributions separately, denoting the symmetries of the vestigial orders as a product of a spatial part $\Phi_\Gamma$, a bilinear in $\bm\varphi$, and spin part $ D_{\Gamma_s}$, a bilinear in $\bar{d}$, via $ V[\Delta^*,\Delta]= \Phi_\Gamma \otimes D_{\Gamma_s}$. In each case the irreps are labelled by $\Gamma=\{A_{1g}, A_{2g}, E_{2g}, E_{1u}\}$ and $\Gamma_s=\{\text{scalar, axial-vector}\}$. Table \ref{tab:vestigials} lists the structure of the spatial and spin vestigials; their connection to applied probes is shown in Fig. \ref{fig1}b.

Our theoretical scenario is illustrated in Fig. \ref{fig1}b: with the application of strain or fields, a particular vestigial combination is induced out of the melted order parameter manifold, producing a ground state expectation value which depends on the nature of the external probe. To determine the precise response to external perturbations, we write down the coupling of the excitonic order parameter with strain and electromagnetic fields via the Landau-Ginzburg approach,
\begin{align}
\label{F}
{\cal F} = {\cal F}_0[\Delta] +\delta {\cal F}[\Delta; \bm E, \bm B, \bm \epsilon]
\end{align}
where ${\cal F}_0[\Delta]$ is a function of the excitonic order parameter, while $\delta {\cal F}[\Delta; \bm E, \bm B, \bm \epsilon]$ contains the allowed couplings to strain $\bm \epsilon$ and electric/magnetic fields; we present the full expressions in the SM.  Via minimisation of the free energy subject to various combinations of external fields, our task is to confirm: (1) the existence of vestigial order $\langle V[\Delta^*,\Delta] \rangle \neq 0$, and (2) that these vestigials are able to explain the array of observed responses near $T_*$. 

\subsection{Unconventional responses}

\subsubsection{Large induced Hall effect and absence of Kerr rotation}

At zero magnetic field no anomalous Hall effect (AHE) is observed experimentally, however a large Hall conductance is induced by an applied field, which is observed to be nonlinear in $B$ \cite{yang2021giant}.

In the absence of strain, the ground state of ${\cal F}_0$, generically favours chiral spatial order $\Phi_{A_{2g}} = i \bm{\varphi}^* \times \bm{\varphi}$, since this nodeless order parameter maximises the condensate energy. This bilinear transforms as an out-of-plane Ising magnetisation, and breaks TRS and mirror symmetries. Yet, in the presence of strain, the favoured bilinear is a rotational symmetry breaking nematic state $\Phi_{E_{2g}} = |\varphi_x|^2- |\varphi_y|^2$. Moreover, in the presence of strain and $B_z$, our theory predicts a piezomagnetic response (see Sec. \ref{piez}), which generates a magnetization that is nonlinear in $B_z$, as shown in Fig.~\ref{fig2}a. Hence, our framework explains the results of Ref.~\cite{yang2021giant} by suggesting that their samples contained an intrinsic or extrinsic strain field, a common occurrence in kagome metals.

The lack of zero-field AHE implies the lack of Kerr rotation, consistent with experiment \cite{saykin2023high}. Conversely, our arguments suggest that ultra low strain samples should in fact exhibit zero field AHE and Kerr rotation.

\subsubsection{eMChA}

Refs. \cite{Guo2022, emcha_Guo2024} observe an anisotropic magnetoelectric contribution to the resistance, $R(\bm{B},\bm{I}) = R_0(1 + \mu^2 B^2 + \gamma_{ij} B_i I_j)$. The term $\mu^2 B^2$, where $\mu$ is the carrier mobility, is the ordinary magnetoresistance of non-magnetic metals, whereas the term proportional to the tensor $\gamma_{ij}$ can only be nonzero when inversion symmetry is broken. The experimental setup takes $\bm B=B \hat{x}$ and $\bm I = I \hat{z}$, and finds that a small out-of-plane $B_z$ changes the sign of $\gamma_{ij}$. We find that these external perturbations couple to the exciton vestigial $\gamma_{ij} \propto  \text{Re}\left[(\Phi_{E_{1u}})_j\otimes ( D_\text{axial})_i\right]$, triggering a transition out of the $A_{2g}$ state. Computing the eMChA tensor from the free energy \eqref{F}, we find a nonzero response with $B_z$ (Fig. \ref{fig2}b).

\subsubsection{Piezomagnetism}
\label{piez}
STM \cite{piezo_xing2024} and transport \cite{anisotropic_Guo2024} report a piezomagnetic response tensor, $P_{ijk}$, such that $\varepsilon_{xy} = P_{xyz} B_z \in E_{2g}$, i.e. produces a strain that changes sign with $B_z$. In our theory, strain induces $\hat{P} \propto \Phi_{E_{2g}} \otimes D''_\text{axial}$ {\color{red}}; our computed piezomagnetic response changes sign with $B_z$, Fig. \ref{fig2}d. Physically, the nematic vestigial can be favoured either via small extrinsic strains of the lattice or via coupling to a $C_3$-breaking CDW.

\subsubsection{Electrostriction}

Ref. \cite{piezo_xing2024} also measures an electro-striction response, i.e. an electric field induced by applied strain. The electro-striction response tensor is $\chi_{ijkl}$, whereby $\varepsilon_{xy} = \chi_{xyxy} E_x E_y \in E_{2g}$. Already $\varepsilon_{xy}  E_x E_y \in A_{1g} + E_{2g}$ and hence is allowed simply by the crystal symmetries without recourse to an order parameter. However, we note that the $E_{2g}$ part of $\varepsilon_{xy}  E_x E_y$ can couple to the $E_{2g}$ vestigial, and hence the onset of vestigial order is expected to enhance this response.

\subsubsection{Elastoresistance}

Experiments \cite{elasto_liu2023, elasto_asaba2023} report a highly unusual elastoresistance response: upon applying strain, one measures $\Delta R_{A_{1g}} = R_{xx} + R_{yy}$ and $\Delta R_{E_{2g}}= R_{xx}-R_{yy}$, and finds a large response in the $A_{1g}$ channel with negligible contribution in the $E_{2g}$ channel -- which naively seems to be at odds with the observation of large strain-induced transport anisotropy \cite{anisotropic_Guo2024}, yet the peak in $A_{1g}$ is further evidence of a phase transition occurring at $T_*$. While Ref. \cite{elasto_liu2023} claims an $E_{2g}$ response is absent, there is clearly a feature at $T_*$, especially in the transverse technique \footnote{Certain elastoresistance measurement geometries suffer from contamination -- an erroneous response in $E_{2g}$ can occur due to some mixing with $A_{1g}$. Importantly, the {\it transverse technique} is designed to be free of contamination, and hence we treat it as reliable evidence of a weak $E_{2g}$ signal.}.

Interestingly, the vestigial $E_{2g} = (E_{1u})^2$ transforms with an effective angular momentum of $\ell=10$ (see the SM), while strain couples at leading order with $\ell=2$. This results in a suppression of the coupling between the $E_{2g}$ and strain in elastoresistance. To demonstrate this, first note the underlying exciton takes the functional form
    \begin{align}
    \label{exciton_l10}
        \bm \varphi\cdot {\cal E}_{\bm k} = (X_{\bm k}^2-Y_{\bm k}^2)Y_{\bm k}(3X_{\bm k}^2-Y_{\bm k}^2)
    \end{align}
The bilinear
    \begin{align}
        \Phi_{\bm k} \sim (\bm \varphi\cdot {\cal E}_{\bm k})^2 = \Phi^{A_{1g}}_{\bm k} + \Phi^{E_{2g}}_{\bm k} ,
    \end{align}
is a mixture of $A_{1g}$ and $E_{2g}$ in the $\ell=10$ harmonic. At leading order strain transforms as
$\epsilon_{\bm k}^{A_{1g}} \sim X_{\bm k}^2+Y_{\bm k}^2$, $\epsilon_{\bm k}^{E_{2g}} \sim X_{\bm k}^2-Y_{\bm k}^2$; and using that resistive anisotropy is linearly proportional electronic anisotropy (see e.g. Ref. \cite{BirolResistivity}) we compute the overlap
\begin{align}
    \Delta R_{A_{1g}} \sim \sum_{\bm k} \Phi_{\bm k} \epsilon_{\bm k}^{A_{1g}},\ \ \ \Delta R_{E_{2g}} \sim \sum_{\bm k} \Phi_{\bm k} \epsilon_{\bm k}^{E_{2g}}.
\end{align}
We find that $\Delta R_{A_{1g}}/\Delta R_{E_{2g}}\approx 15$. We suggest that this relative suppression of $\Delta R_{E_{2g}}$ explains the large peak in the $A_{1g}$ susceptibility alongside a small ratio of the $E_{2g}$ and $A_{1g}$  responses.

\subsubsection{Superconducting diode effect}

Recently a zero-field superconducting diode effect has been observed \cite{le2024superconducting}. Such an effect requires broken TRS, $I$ and $C_{2z}$ \cite{Scammell_Diode_2022, Lin2022}; crucially, inversion symmetry must be broken in the normal state for this effect to arise. The excitonic vestigial $\bm{\Phi}_{E_{1u}}$ is sufficient to break $I$ and $C_{2z}$, satisfying the symmetry requirements for a superconducting diode effect.


\subsubsection{Raman spectroscopy}

Below $T_*$, Raman spectroscopy should observe new $A_{1g}, E_{2g}$ modes directly, as well as the appearance of a peak due to the $E_{1u}$ phonon which is Raman inactive above $T_*$, but becomes visible due to the coupling with $E_{1u}$ excitons. 


The Raman probe $J_{XX}$ couples the $E_{1u}$ phonon doublet $\boldsymbol{O}_{E_{1u}}=\left(O_X, O_Y\right)$ to the $E_{1u}$ order via $J_{X X} \boldsymbol{O}_{E_{1u}} \cdot \boldsymbol{\Phi}_{E_{1u}}$.  Defining axes such condensation occurs with $\Phi_X>0$ and $\Phi_Y=0$, implies that the phonon doublet $\left(O_X, O_Y\right)$ is split, with $O_X$ now possessing lower excitation energy due to coupling with $\Phi_X$. Rotating the polarisation of incident light, $J_{R R}^\phi=\cos ^2 \phi J_{X X}+\sin ^2 \phi J_{Y Y}$; we therefore expect a dominant contribution
\begin{gather}
J_{R R}^\phi \boldsymbol{O}_{E_1} \cdot \boldsymbol{\Phi}_{E_1}
=\cos ^2 \phi J_{X X} O_X \Phi_X.
\end{gather}
Hence, \textit{one should observe a Raman peak with a twofold rotational dependence in $\phi$}. Recent studies \cite{Jin_PRL_2024} and unpublished data \cite{Note8} have observed such a mode with two-fold dependence that sharpens near $T_*$ \cite{Note9}.

\section{Discussion}

\subsection{Subsidiary CDW orders}

\begin{figure}[t!]
  \begin{center}
    \includegraphics[width=0.85\linewidth]{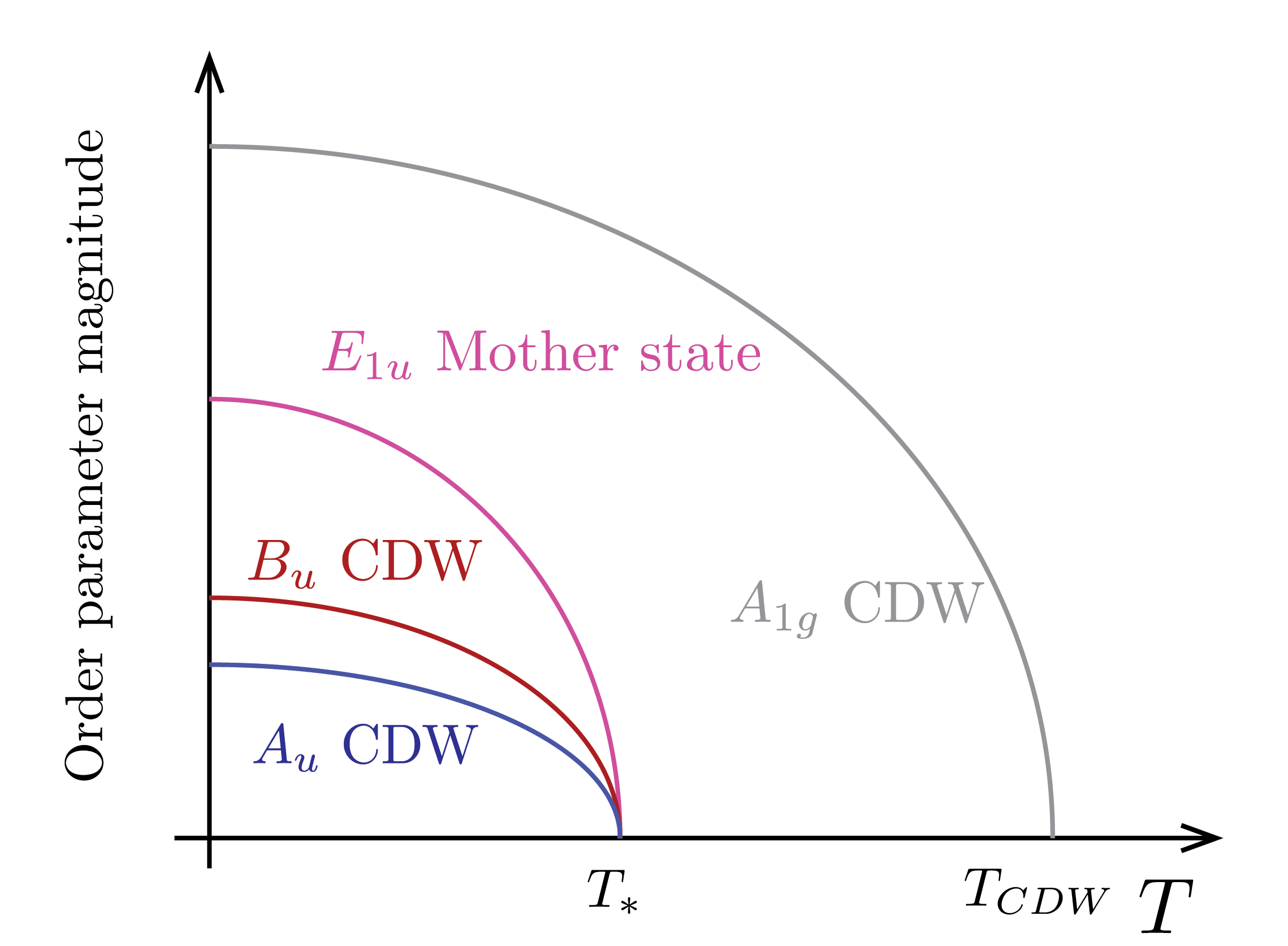}
    \caption{\textbf{Induced CDW orders.} The proposed symmetry breaking pattern: $A_g$ CDW at appears at $T_{CDW}$ followed by vestigial mother state at $T_*$, here taken to be $\Phi_{E_{1u}}$, which necessarily admixes $A_u, B_u$ CDWs.}\label{fig3}
	\end{center}
    \vspace{-0.3cm}
\end{figure}

A final interesting effect we discuss is the induction of subsidiary CDW orders via the excitonic vestigial state. We have claimed that in zero external fields the dominant vestigial is $\bm \Phi_{E_{1u}} D_\text{scalar}$; we set $D_\text{scalar}=1$ for clarity below. Denoting a charge density wave order with wavevector $\bm{M}_i$ transforming with symmetry irrep $\Gamma$ (w.r.t. little group $C_{2h}$ at the $\bm M$ point) as $\mathcal{C}^{\Gamma}_{\bm{M}_i}$, the allowed couplings to CDW orders are 
\begin{gather}
\notag {\cal F}_\text{mix} = \left[\alpha_1 \ \left({\cal C}_{\bm M_i}^{A_g} \bm V_{ij} {\cal C}_{\bm M_j}^{A_u}\right)+\alpha_2 \ \left({\cal C}_{\bm M_i}^{A_g} \bm V_{ij} {\cal C}_{\bm M_j}^{B_u}\right)\right]\cdot \bm \Phi_{E_{1u}}
\end{gather}
 where $\bm V_{ij}$ is a doublet of $3\times3$ matrices $V^{(1)}=$ diag$(1,-2,1)$ and $V^{(2)}=$ diag$(1,0,-1)$. Here we have labeled the CDWs via the little group at $M$-points, $C_{2h}$.  It follows that $B_u,A_u$ CDW orders become admixed via the $E_{1u}$ exciton vestigial \cite{Note10} -- the resulting phase diagram is sketched in Fig. \ref{fig3}. One concrete prediction of this scenario is the hierarchy of order parameter magnitudes illustrated by the vertical axis.

\subsection{Conclusions and outlook}

Our work produces a unified organizational principle to reconcile the diverse and seemingly contradictory experimental responses observed in kagome metals. At its core is a fluctuation-melted excitonic mother state, whose vestigial orders hosts a range of possible symmetry-breaking patterns, which emerge or are enhanced under the application of external fields.

To demonstrate the utility of this framework, we employed explicit modeling via a Landau free energy treatment to account for the complex phenomenology observed across various experiments on kagome metal CsV$_3$Sb$_5$. Taken together, our findings provide strong evidence that vestigial orders explain the puzzling observations near $T_*$, challenging the prevailing view that CDW order alone underlies the rich phenomenology observed below $T_\text{CDW}$, including $T_*$.


Prospectively, we suggest experiments aimed at directly probing the proposed excitonic mother state:
\begin{itemize}
\item[(i)] Optical spectroscopy: The evolution of excitonic bound-state spectra in the range $T_*<T<T_\text{CDW}$ could provide evidence for or against the formation of the excitonic mother state. If the excitation energy vanishes as $T_*$ is approached, this would be consistent with an excitonic mother state.
\item[(ii)] Scanning tunneling microscopy (STM): Since STM is sensitive to spatial symmetries, a detailed analysis near $T_*$ would be particularly informative -- in particular a comparison of the magnitude of $E_{1u}$ and CDW $A_u$/$B_u$ order parameter magnitudes, c.f. Fig. \ref{fig3}.
\item[(iii)] Symmetry-selective ultrasound experiments: By measuring the system's elastic moduli, these experiments can effectively uncover the point-group symmetries of order parameters \cite{Ghosh_Ramshaw_SciAdv_2020, Theuss2024}.
\end{itemize}

The possible destruction of the excitonic mother state with pressure or doping may account for abrupt changes in the superconducting and magnetic properties observed in $A$V$_3$Sb$_5$; exploring the influence of such perturbations on the responses discussed above will yield important insights into the nature of this unconventional mother state.

\section*{Acknowledgements}

We thank Riccardo Comin, Chenhao Jin, and Dongjin Oh for helpful discussions and comments. J.I. is supported by NSF Career Award No. DMR-2340394.

\bibliography{refs.bib}

\widetext

\appendix

\section{Effective two-orbital tight binding model}
There are three vHS in vicinity of the $M$-points, these have orbital character $d_{zx}, d_{yz}, d_{xy/x^2-y^2}$ \cite{Hu2022, Kang2022}. The vHS of predominant $d_{xy/x^2-y^2}$ character is most strongly reconstructed below $T_\text{CDW}$ \cite{Kang2022}. The m-type vHS, of $d_{yz}$ character, also shows reconstruction, bu it is seen to be weaker and moreover the p-type vHS, of $d_{zx}$ character, is unreconstructed \cite{Kang2022}. Based on these observations, our effective two-band model comprises the just $d_{zx}, d_{yz}$ bands and makes the simplification that $d_{yz}$ is not reconstructed. The corresponding tight binding model was established in \cite{PhysRevLett.127.177001}
\begin{align}
H= & \sum_{\mathbf{k} i \alpha} \epsilon_\alpha \psi_{\mathbf{k} i \alpha}^{\dagger} \psi_{\mathbf{k} i \alpha}-\sum_{\mathbf{k} i j \alpha} t_\alpha \Phi_{i j}(\mathbf{k}) \psi_{\mathbf{k} j \alpha}^{\dagger} \psi_{\mathbf{k} i \alpha}  -t^{\prime} \sum_{\mathbf{k}, i j} \Phi_{i j}(\mathbf{k}) s_{i j}\left(\psi_{\mathbf{k} j \mathrm{xz}}^{\dagger} \psi_{\mathbf{k} i \mathbf{} z}-\psi_{\mathbf{k} j \mathrm{yz}}^{\dagger} \psi_{\mathbf{k} i \mathrm{xz}}\right).
\end{align}
Here $\psi_{\mathbf{k} i \alpha}^{\dagger}$ creates an electron at quasimomentum $\bm k$, sublattice $i=A,B,C$ and in orbital $\alpha= xz, yz$. The lattice hopping factors are $\Phi_{A B}(\mathbf{k})=1+e^{-2 i \mathbf{k} \cdot \mathbf{a}_1}, \Phi_{B C}(\mathbf{k})=1+e^{-2 i \mathbf{k} \cdot \mathbf{a}_3}$, and $\Phi_{A C}(\mathbf{k})=1+e^{-2 i \mathbf{k} \cdot \mathbf{a}_2}$ and obey condition $\Phi_{i j}(\mathbf{k})=\left(1-\delta_{i j}\right) \Phi_{j i}^*(\mathbf{k})$. Here wde have denoted by $\mathbf{a}_{1,2}=(\sqrt{3} / 2, \pm 1 / 2)^T$ and $\mathbf{a}_3=$ $(0,-1)^T$ the vectors connecting sublattices. The parameters $\epsilon_\alpha, t_\alpha, t'$ correspond to onsite orbital energy, orbital dependent nearest-neighbour hopping, and an interorbital coupling. Finally, $s_{ij}$ accounts for the non-trivial spatial symmetry of the interorbital overlap of $d_{zx}, d_{yz}$, i.e. $s_{ij}\in B_{2u}$. Explicitly, assuming the sublattice sites are labeled according to \cite{PhysRevLett.127.177001}, then $s_{AC}=s_{CB}=-s_{AB}$ and $s_{ij}=-s_{ji}$.

\section{Meanfield gap equation and $E_{1u}$ excitonic order}
We briefly outline the origin of the $E_{1u}$ excitonic order. The $E_{1u}$ follows from a direct mean field theory or patch theory treatment (detailed below) and can also be motivated more heuristically as follows. Firstly, a mean field and/or patch theory favours $E_{2g}$ pairing of the BdG gap function since this provides a full gapping of the Fermi surface and thereby maximises the condensate energy. Second, further account of orbital form-factors introduces an additional $B_{2u}$ -- via reasoning detailed below -- and the overall irrep is therefore $E_{2g}\times B_{2u}=E_{1u}$. To account for the $B_{2u}$ orbital form factors, we note that the two vHS exhibit $m$- and $p$-type sublattice support, and in the case of CsV$_3$Sb$_5$, for $T<T_\text{CDW}$ the $m$-type vHS are primarily composed of the $d_{zx}$ orbitals of the vanadium kagome lattice, while the (un-reconstructed) $p$-type vHS arises from a band composed of the $d_{yz}$ orbitals (noting that the $p$-type vHS arising from the $d_{z^2}/d_{x^2-y^2}$ orbitals undergoes significant reconstruction in the CDW phase \cite{Kang2022, Hu2022}). Since the $p$ ($m$)-type wavefunctions are inversion even (odd), the product of the band irreps of the TvHS -- relevant for excitonic pairing -- falls into $B_{2u}$ of $D_{6h}$.
 
Here we sketch the meanfield gap equation. Decoupling specifically in the excitonic spin-singlet channel, the resulting ``BdG'' mean-field Hamiltonian is 
\begin{align}
    H_\text{BdG}&=\begin{pmatrix} c^
    \dag_{\bm k} \\ v^\dag_{\bm k}\end{pmatrix}^T\begin{pmatrix} \varepsilon^c_{\bm k} & M_{\bm k} \\ M^*_{\bm k} & \varepsilon^v_{\bm k}\end{pmatrix}\begin{pmatrix} c^
    \dag_{\bm k} \\ v^\dag_{\bm k}\end{pmatrix}=\begin{pmatrix} \tilde{c}^
    \dag_{\bm k} \\ \tilde{v}^\dag_{\bm k}\end{pmatrix}^T\begin{pmatrix} \varepsilon^c_{\bm k} & M_{\bm k}\chi^{B_{2u}}_{\bm k} \\ M^*_{\bm k}\chi^{B_{2u}}_{\bm k} & \varepsilon^v_{\bm k}\end{pmatrix}\begin{pmatrix} \tilde{c}^
    \dag_{\bm k} \\ \tilde{v}^\dag_{\bm k}\end{pmatrix}.
\end{align}
In the second equality, we have used that $(\chi^{B_{2u}}_{\bm k})^2=\chi^{A_{1g}}_{\bm k}$ and further simplified by taking $\chi^{A_{1g}}\approx 1$. 

An interaction term, in the excitonic channel, then rescales as
\begin{align}
V_0 c_{\bm k_1}^\dag v_{\bm k_1} v_{\bm k_2}^\dag c_{\bm k_2} \to V_0 (\chi^{B_{2u}}_{\bm k_1} \chi^{B_{2u}}_{\bm k_2})^2 \tilde{c}_{\bm k_1}^\dag \tilde{v}_{\bm k_1} \tilde{v}_{\bm k_2}^\dag \tilde{c}_{\bm k_2}=\tilde{V}_{0;\bm k_1,\bm k_2} \tilde{c}_{\bm k_1}^\dag \tilde{v}_{\bm k_1} \tilde{v}_{\bm k_2}^\dag \tilde{c}_{\bm k_2}.
\end{align}
Following the standard procedure, we obtain the meanfield gap equation in the order parameter $\tilde{M}_{\bm k}\equiv M_{\bm k}\chi^{B_{2u}}_{\bm k}$,
\begin{align}
    \tilde{M}_{\bm k_1} & = \frac{1}{2}\sum_{\bm k_2} \tilde{V}_{0;\bm k_1,\bm k_2}\frac{\tanh\left(\frac{\varepsilon_{\bm k_2}^c}{2T}\right)-\tanh\left(\frac{\varepsilon_{\bm k_2}^c}{2T}\right)}{\varepsilon_{\bm k_2}^c-\varepsilon_{\bm k_2}^v} \tilde{M}_{\bm k_2}.
\end{align}
Previous work \cite{ingham2024theory} has shown the conditions for the interactions (here denoted $\tilde{V}_{0;\bm k_1,\bm k_2}$) to produce $E_{2g}$ ordering. We therefore do not repeat those lengthy calculations here. Instead, using those results, we infer that a candidate (leading) solution of the gap equation has $\tilde{M}_{\bm k} \in E_{2g}$, and hence $M_{\bm k} \in B_{2u}\otimes E_{2g} = E_{1u}$. 

In terms of basis functions, $B_{2u} = Y_{\bm k}(3X_{\bm k}^2-Y_{\bm k}^2)$ while $E_{2g}=\{(X_{\bm k}^2-Y_{\bm k}^2),2X_{\bm k} Y_{\bm k}\}$. This suggests that the $E_{1u}$ order for this system is
${\cal E}_{\bm k}^{E_{1u}}= \{(X_{\bm k}^2-Y_{\bm k}^2),2X_{\bm k} Y_{\bm k}\} Y_{\bm k}(3X_{\bm k}^2-Y_{\bm k}^2)$, i.e. excitonic pairing has angular momentum $|\ell|=5$. 

\section{Free energy and coupling to external fields}
We consider exciton pairing in a 2D irrep, here explicitly $E_{1u}$ of $D_{6h}$, and further taking the approximate $\text{SU}(2)_{+} \times \text{SU}(2)_{-}$ spin symmetry to be exact we obtain the allowed terms in the free energy expansion. To begin we define the pairing function $M_{\bm k}$ of Eq. \eqref{Hpair} as 
\begin{align}
M_{\bm k}=\sum_{\mu= \pm}\left(X_{\bm k}+i \mu Y_{\bm k}\right) \Delta_\mu, \quad \Delta_\mu=\sum_{\nu=0}^3 \eta_{\mu \nu} \sigma_\nu,
\end{align}
and obtain the action of the symmetries
\begin{align}
\notag C_{3z}:\left(\Delta_{+}, \Delta_{-}\right) & \longrightarrow\left(\omega \Delta_{+}, \omega^* \Delta_{-}\right), \\
\notag C_{2z}:\left(\Delta_{+}, \Delta_{-}\right) & \longrightarrow\left(- \Delta_{+}, - \Delta_{-}\right), \\
\notag C_{2x}:\left(\Delta_{+}, \Delta_{-}\right) & \longrightarrow\left(\Delta_{-}, \Delta_{+}\right), \\
\notag \sigma_{h}:\left(\Delta_{+}, \Delta_{-}\right) & \longrightarrow\left(\Delta_{+}, \Delta_{-}\right), \\
\notag \Theta:\left(\Delta_{+}, \Delta_{-}\right) \notag & \longrightarrow\left(\Delta_{-}^{\dagger}, \Delta_{+}^{\dagger}\right), \\
\notag \mathrm{SU}(2)_{+} \times \mathrm{SU}(2)_{-}: \Delta_\mu & \longrightarrow e^{-i \gamma_{+} \cdot \sigma} \Delta_\mu e^{i \gamma_{-} \cdot \sigma}, \\
\mathrm{U}(1): \Delta_\mu & \longrightarrow e^{i \theta} \Delta_\mu.
\end{align}
Note that the action of the point group symmetries depends on our choice of irrep, which we here take to be $E_{1u}$.
Armed with these transformations, all allowed terms in the free energy, up to quartic order, are
\begin{align}
\notag \mathcal{F}_0 &=  a \sum_{\mu= \pm} \operatorname{tr}\left[\Delta_\mu^{\dagger} \Delta_\mu\right]+\frac{b_1}{4}\left(\sum_{\mu= \pm} \operatorname{tr}\left[\Delta_\mu^{\dagger} \Delta_\mu\right]\right)^2 \\
\notag &+\frac{b_2}{2} \sum_{\mu= \pm} \operatorname{tr}\left[\Delta_\mu^{\dagger} \Delta_\mu \Delta_\mu^{\dagger} \Delta_\mu\right] \\
\notag & +\frac{b_3}{4} \operatorname{tr}\left[\Delta_{+}^{\dagger} \Delta_{+}\right] \operatorname{tr}\left[\Delta_{-}^{\dagger} \Delta_{-}\right]+\frac{b_4}{4}\left|\operatorname{tr}\left[\Delta_{+}^{\dagger} \Delta_{-}\right]\right|^2 \\
\notag & +\frac{b_5}{2}\left(\operatorname{tr}\left[\Delta_{+}^{\dagger} \Delta_{+} \Delta_{-}^{\dagger} \Delta_{-}\right]+\operatorname{tr}\left[\Delta_{-} \Delta_{-}^{\dagger} \Delta_{+} \Delta_{+}^{\dagger}\right]\right),\\
{\cal F}_0' &= c_1 \sum_{i=x,y}\sum_{\nu=0,x,y,z}\text{Re}(\varphi_0^*\varphi_i d_\nu)^2 + c_2 c_2 \delta_{\nu,0}|d_\nu\varphi|^2
\end{align}
Here ${\cal F}_0'$ accounts for explicit breaking of global symmetries: term with coupling constant $c_1$ accounts for U(1)-breaking, i.e. it contains the background field $\varphi_0$, while the term coupling constant $c_2$ accounts for SO(4)$\to$SO(3) due to spin-exchange interactions. 

We consider external probes $E_i, B_i, \epsilon_{ij}$ and find how they couple to the vestigials (i.e. bilinears in $\Delta_\mu$) in the free energy. We consider only the bilinear in $\Delta_\mu$ and up to second order in external fields, and only keep non-trivial terms (i.e. the external fields are in a non-trivial irrep). The external fields couple via
\begin{align}
    \notag &\delta {\cal F}[\bm B]= g_{1}^B B_z \Phi_{A_{2g}}(\bar{d}^*\cdot\bar{d})-g_{2}^B \Phi_{A_{1g}}2\text{Im}(d_0^* \bm d)\cdot \bm B+ g_{3}^B B_z\Phi_{A_{2g}}2\text{Im}(d_0^* \bm d)\cdot \bm B+g_{4}^B(B_x^2-B_y^2,2B_xB_y)\cdot\bm \Phi_{E_{2g}}(\bar{d}^*\cdot\bar{d}),\\
    \notag &\delta {\cal F}[\bm E]= g_{1}^E 2\text{Re}(\bm E\cdot \bm\varphi d_0\varphi_0^*)+ g_{2}^E (E_x^2-E_y^2,2E_xE_y)\cdot\bm \Phi_{E_{2g}}(\bar{d}^*\cdot\bar{d}),\\
    \notag &\delta {\cal F}[\bm\epsilon]= g^\epsilon \bm\epsilon\cdot\bm \Phi_{E_{2g}}(\bar{d}^*\cdot\bar{d}),\\
    \notag  &\delta {\cal F}[\bm B,\bm \epsilon]= g^{\epsilon B} \bm \epsilon\cdot \bm \Phi_{E_{2g}}2\text{Im}(d_0^* \bm d)\cdot \bm B,\\
    \notag &\delta {\cal F}[\bm E,\bm \epsilon]= g_1^{\epsilon E}\bm \epsilon \cdot (E_x^2-E_y^2, 2E_x E_y)\Phi_{A_{1g}}(\bar{d}^*\cdot\bar{d}) +g_{2}^{\epsilon E}(\epsilon_1\Omega_1-\epsilon_2 \Omega_2,\epsilon_1\Omega_2+\epsilon_2 \Omega_1)\cdot \Phi_{E_{2g}}(\bar{d}^*\cdot\bar{d}),\\
    \notag &\delta {\cal F}[\bm E,\bm B]= \sum_{i=x,y} \sum_{j=x,y,z}g_{ij}^{EB} E_i B_j 2\text{Im}(\varphi_0^*\varphi_i d_j).
\end{align}
Along the way we introduced $\{\Omega_1,\Omega_2\}\equiv\{E_x^2-E_y^2, 2E_xE_y\}$.   Comment: imposing that the conduction and valence bands have a common spin $g$-factor, one can show that the axial $\bm d^*\times \bm d$ does not couple to external magnetic field, yet $\bm{d}$ does. The precise values of the coefficients in the free energy can be computed microscopically by integrating out the electrons in a mean-field treatment (see the Supplemental Materials of e.g. Refs. \cite{li2020artificial, ingham2024theory, scammell2022intrinsic, ingham2023quadratic, guerci2024topological} for illustration); we leave this calculation to future work.\\

In the main text Figure \ref{fig2}, the following coefficients were employed: $g^{EB}_{ij}=0.5, g^B_{i}=g^E_{i'}=g^\epsilon=g_{i''}^{\epsilon E}=g^{\epsilon B}=1$. (a)
$\epsilon_1=-\epsilon_2=\{0,0.2,0.4,0.6\}$
$a=-0.05, b_i=1, c_1=0, c_2=-0.05$. 
(b)
$E_z=B_y=1$
$\epsilon_1=-\epsilon_2=0.15$
$a=-0.05, b_i=1, c_1=0, c_2=-0.05$. 
(c)
$\epsilon_1=-\epsilon_2=0.15$
$a=-0.05, b_i=1, c_1=0, c_2=-0.05$. 
(d)
$\epsilon_1=0.25,\epsilon_2=0$,
$a=(T-T_c), T_c=1, b_i=1, c_1=0, c_2=-0.05$.

\end{document}